\documentstyle[12pt]{article}

\input epsf

\newcommand{\be}{\begin{equation}}
\newcommand{\ee}{\end{equation}}
\newcommand{\bea}{\begin{eqnarray}}
\newcommand{\eea}{\end{eqnarray}}
\newcommand{\ba}{\begin{array}}
\newcommand{\ea}{\end{array}}
\newcommand{\non}{\nonumber}
\newcommand{\bm}[1]{\mbox{\boldmath $#1$}}
\newcommand{\RRR}{{\bm R}}

\newcommand{\ol}{\overline}
\newcommand{\rrr}[2]{\rule[-#1pt]{0pt}{#2pt}}
\newcommand{\hsp}[1]{\hspace*{#1pt}}
\newcommand{\subst}[3]
{${\varrho_{#1}}: \ba{r@{\:}c@{\:}l} 
 a & \rightarrow & #2 \\         
 b & \rightarrow & #3 \ea$}

\newcommand{\EPL}[3]{Europhys.\ Lett.\ {\bf #1} (19#2) #3}

\newcommand{\IJMPB}[3]{Int.\ J.\ Mod.\ Phys.\ {\bf B#1} (19#2) #3}
\newcommand{\JPhysA}[3]{J.\ Phys.\ {\bf A#1} (19#2) #3}

\newcommand{\JSP}[3]{J.\ Stat.\ Phys.\ {\bf #1} (19#2) #3}

\newcommand{\NPB}[3]{Nucl.\ Phys.\ {\bf B#1} (19#2) #3}

\newcommand{\PhysRev}[3]{Phys.\ Rev.\ {\bf #1} (19#2) #3}

\newcommand{\RevMP}[3]{Rev.\ Math.\ Phys.\ {\bf #1} (19#2) #3}
\newcommand{\ZPhys}[3]{Z.\ Phys.\ {\bf #1} (19#2) #3}

\renewcommand{\baselinestretch}{1.1}

\begin{document}


\begin{center}
{\LARGE\bf Non-Periodic Ising Quantum Chains \\[3mm]
and Conformal Invariance} \\[8mm]
{\large\sc Uwe Grimm$^{1,3}$\footnotetext[3]{Address after 
           1.9.1993: Instituut voor Theoretische Fysica, Universiteit van
            Amsterdam, \newline Valckenierstraat 65, 1018 XE Amsterdam, The
            Netherlands} \hsp{1} and \hsp{1} Michael Baake$^{2}$} \\[4mm]
{\footnotesize \mbox{}\footnotemark[1]
 Department of Mathematics, University of Melbourne,\\
        Parkville, Victoria 3052, Australia} \\[2mm]
{\footnotesize \mbox{}\footnotemark[2]
 Institut f\"{u}r Theoretische Physik, Universit\"{a}t T\"{u}bingen,\\
        Auf der Morgenstelle 14, 72076 T\"{u}bingen, Germany} \\[6mm]
July 1993 \\[6mm]
\end{center}


\begin{quote}
{\small\sf 
In a recent paper, Luck \cite{Luck} investigated the critical behaviour
of one-dimensional Ising quantum chains with couplings constants
modulated according to general non-periodic sequences.
In this short note, we take a closer look at the case
where the sequences are obtained from 
(two-letter) substitution rules and
at the consequences of Luck's results at criticality. 
They imply that only for a certain class of substitution rules 
the long-distance behaviour is still
described by the c=1/2 conformal field theory of a free Majorana 
fermion as for the periodic Ising quantum chain, whereas the general case
does not lead to a conformally invariant scaling limit.}

{\small\bf Key Words:} \hsp{5} 
{\footnotesize 
 quantum \hfill spin \hfill chains, \hfill 
 non-periodic \hfill systems, \hfill
 substitution \hfill rules, \hfill conformal \\
 \hsp{77} invariance, \hsp{1} Pisot-Vijayaraghavan numbers}
\end{quote}



\renewcommand{\theequation}{\arabic{section}.\arabic{equation}}

%
%
%
%

\section{Introduction}
\setcounter{equation}{0}

In this note, we consider the Ising quantum chain
with ferromagnetic exchange couplings $\varepsilon_{j}>0$
which follow an, in general, non-periodic sequence of 
finitely many different values.
For convenience, we choose the (constant) transversal 
field to be equal to one,
and consider the following Hamiltonians
\be
 H^{(\pm)} \;\; = \;\; 
-\,\frac{1}{2}\;\left(\;\sum_{j=1}^{N}\:
\varepsilon_{j}\,\sigma_{j}^{x}\,\sigma_{j+1}^{x} 
\;\; +\;\;\sum_{j=1}^{N}\,\sigma_{j}^{z}\right) \label{H}
\ee
of (anti-) periodic approximants ($\sigma_{N+1}^{x}=\pm\sigma_{1}^{x}$)
obtained by truncating the sequence of exchange couplings to
the first $N$  elements. Here, $\sigma_{j}^{\alpha}$ denotes the
Pauli matrix $\sigma^{\alpha}$ acting on site $j$ of an $N$-fold
tensor product space.
Note that $H^{(\pm)}$ commutes with the operator
\be
Q \;\; = \;\; \prod_{j=1}^{N} \sigma_{j}^{z} 
\; , \hsp{20} Q^2 \;\; = \;\;  Id \; ,\label{Q}
\ee
which has eigenvalues $\pm 1$. 
We denote the projectors onto the corresponding eigenspaces 
(sometimes also called sectors) by
\be
P_{\pm} \;\; = \;\; \frac{1}{2}\: (Id \pm Q)\; .
\label{P} 
\ee

In ref.~\cite{Luck},
the Hamiltonian (\ref{H}) is formulated in terms
of fermionic operators by means of a Jordan-Wigner
transformation \cite{Lieb} and diagonalized by
a suitable Bogoljubov-Valatin transformation. 
To be more precise, it is
in fact the so-called mixed sector Hamiltonian \cite{BCS} 
$\tilde{H}^{(+)}$ defined by
\be
\tilde{H}^{(\pm)} \;\; = \;\; H^{(\pm)} P_{+}\: 
+\: H^{(\mp)} P_{-} \label{mix}
\ee
which is considered.
This is necessary since the Jordan-Wigner transformation 
of $H^{(\pm)}$ produces non-local boundary terms in 
the fermionic operators
whereas the above Hamiltonians (\ref{mix}) corresponds to
periodic (resp.\ antiperiodic) boundary conditions in 
terms of the fermions.

Let us cite three results of ref.~\cite{Luck}
which are central to our subsequent discussion.
The system described by the Hamiltonian (\ref{H}) resp.\ (\ref{mix}) is critical 
at $\mu=0$ with
\be 
\mu \;\; = \;\; \lim_{N\rightarrow\infty} \:
\frac{1}{N}\sum_{j=1}^{N}\log(\varepsilon_{j})\; . \label{mu}
\ee
The dispersion relation for the low-energy one-particle 
excitations with energy $\Lambda$ takes the 
following form 
\be
\Lambda^2\;\; =\;\; v^2\: ( q^2 + \mu^2)\; ,
\ee
where $q$ denotes the momentum, and $v$,
the velocity of elementary excitations, is given by 
\be
\frac{1}{v^2}\;\; =\;\; \lim_{N\rightarrow\infty}\:
\frac{1}{N^2} \sum_{j=1}^{N}\sum_{k=1}^{N}\prod_{\ell=1}^{k}
\varepsilon_{j+\ell-1}^2\; , \label{v}
\ee
provided the latter limit exists.

In what follows, we consider sequences of coupling
constants which are obtained by substitution rules,
focussing on the case of two-letter substitution rules
(although most properties can be generalized quite easily
to the $n$-letter case).

%
%
%
%

\section{Substitution Rules and Matrices}
\setcounter{equation}{0}

We consider two-letter substitution rules
\be
{\varrho} : \;\;
\ba{ccc} a & \rightarrow & w_a \\
         b & \rightarrow & w_b \ea \label{subs}
\ee
where $w_{a}$ and $w_{b}$ are words in $a$ and $b$
(we do not allow for inverses of $a$ or $b$ here).
If one defines multiplication of words through
concatenation, the action of $\varrho$ is extended
to arbitrary words in $a$ and $b$ via the homomorphism
property $\varrho(w_a w_b) = \varrho(w_a) \varrho(w_b)$,
for details see ref.~\cite{tracemap} and references included therein.
To $\varrho$ we associate a $2\times 2$ matrix $\RRR_{\varrho}$
\be
\RRR_{\varrho} \;\; =\;\;
\left( \ba{ll}
 \#_a(w_a) & \#_a(w_b) \\
 \#_b(w_a) & \#_b(w_b) \ea \right) \label{submat}
\ee
whose elements count the number of $a$'s and $b$'s 
in the words $w_{a}$ and $w_{b}$, respectively. 
Note that we use the transpose matrix in comparison with
\cite{tracemap} because we need only the {\em statistical} eigenvectors
in our discussion. They are then the right-eigenvectors of
$\RRR_{\varrho}$. With this convention, one also has
$\RRR_{\varrho \circ \sigma} = \RRR_{\varrho} \cdot \RRR_{\sigma}$.
By using the
substitution rule $\varrho$ of (\ref{subs})
iteratively on an initial word $w(0)$, say $w(0)\! =\! a$ for
definiteness, one obtains a sequence of words 
$w(n)=\varrho(w(n\! -\! 1))$. 
Since we are interested in sequences
which have a unique limit word $w$,
we restrict ourselves to substitution rules
where $w_{a}$ begins with the letter $a$.
In this case, the sequence $w(n)=\varrho(w(n-1))$
obviously commences with $w(n\! -\! 1)$ and thus 
each iteration only appends letters to the previous word.
The length (i.e., the number of letters) of the
word $w(n)$ is given by $f_{\varrho}(n)$
defined as follows
\bea
e_{\varrho}(n) & = & 
\left(\ba{l} e^{(a)}_{\varrho}(n) \\ 
e^{(b)}_{\varrho}(n)\ea \right) \;\; = \;\;
\RRR_{\varrho}\: e_{\varrho}(n\! -\! 1) \; ,\hsp{20}
e_{\varrho}(0) \;\; = \;\; \left(\ba{l} 1 \\ 0 \ea\right) \; , \non \\ 
f_{\varrho}(n) & = & e^{(a)}_{\varrho}(n) + e^{(b)}_{\varrho}(n) \; .
\label{length}
\eea

We denote the eigenvalues of $\RRR_{\varrho}$ in (\ref{submat}) by
$\lambda^{(\pm)}_{\varrho}$ where $\lambda^{(+)}_{\varrho}$ stands for
the Perron-Frobenius eigenvalue. The corresponding 
statistically normalized eigenvector is determined by
\be
\RRR_{\varrho} \left( \ba{l} p_{a} \\ p_{b} \ea \right) \;\; = \;\;
\lambda^{(+)}_{\varrho} \left( \ba{l} p_{a} \\ p_{b} \ea \right) \; ,
\hsp{20} p_{a} + p_{b} \;\; =\;\; 1\; .
\ee
The eigenvalue $\lambda^{(+)}_{\varrho}$ determines the asymptotic
inflation factor for one substitution, whereas $p_{a}$ ($p_{b}$)
is the frequency of the letter $a$ (resp. $b$) in
the limit word. Let us now discuss under which conditions $\lambda^{(-)}$ contains 
some information about fluctuations.

\section{Fluctuations}
\setcounter{equation}{0}

Consider the truncated sequences $w^{(N)}$ of the first $N$ letters 
of the limit word $w$ obtained from the substitution rule (\ref{subs})
with initial word $a$. To measure the fluctuation, we define
\be
g(N)\;\; = \;\; \#_{a} (w^{(N)}) - p_{a} N \; ,\hsp{20}
g_{n} \;\; = \;\; g(f_{\varrho}(n))\; , \label{gN}
\ee 
and
\be
h(N) \;\; =\;\; \max_{M\leq N} |g(M)| \; ,\hsp{20}
h_{n} \;\; = \;\; h(f_{\varrho}(n))\; . \label{hN}
\ee
Note that it does not matter whether we look at fluctuations
in the frequency of the letter $a$ or $b$ since
$\#_{a} (w^{(N)}) + \#_{b} (w^{(N)}) = N$ and $p_{a}+p_{b}=1$.
Hence, $g(N)$ just changes sign if one replaces $a$ by $b$ in 
Eqs.~(\ref{gN}) and (\ref{hN}). Of course, 
if $\lambda_{\varrho}^{(+)}$ is not degenerate,
$\lim_{N\rightarrow\infty}(g(N)/N)=0$.

The behaviour of $g(N)$ for words of length $N\! =\! f_{\varrho}(n)$ 
which correspond to proper (or complete)
iteration steps is governed by the second largest
eigenvalue $\lambda^{(-)}$ \cite{Luck}. In fact,
writing the starting vector $e_{\varrho}(0)$ (\ref{length}) as
a linear combination of the eigenvectors of $\RRR_{\varrho}$
(\ref{submat}), one easily verifies $g_{n}\sim{\lambda^{(-)}}^{n}$.
Hence $|\lambda^{(-)}|<1$ implies that
$g_{n}$ converges to zero for $n\rightarrow\infty$.
On the other hand, if $|\lambda^{(-)}|>1$ then
$g_{n}$ in general diverges. In the limiting case of 
$|\lambda^{(-)}|=1$, $|g_{n}|$ is constant 
and therefore is bounded away from zero 
{\em and}\ infinity.
This observation brings along, once again, the concept
of Pisot-Vijayaraghavan numbers (PV-numbers for short) \cite{Cassels}.
They are real algebraic integers $\vartheta >1$ all algebraic conjugates
of which (except $\vartheta$) lie inside the unit circle.
If the characteristic polynomial of $\RRR_{\varrho}$ is irreducible over
the integers and if the Perron-Frobenius eigenvalue is larger than 1, the
PV-property really is what determines the conformal nature of the critical
point. The same seems still to be true if the characteristic polynomial
is reducible but all eigenvalues except the largest one lie inside the
unit circle -- in which case we say that the underlying substitution
has bounded fluctuation property. However, the reducible case
requires some care, as we will demonstrate by an example.

To understand why this is so important, one has to realize that these
considerations only apply to words which are
obtained by proper iteration steps. In between, 
fluctuations can behave quite differently, 
especially in the so-called marginal case $|\lambda^{(-)}|=1$
\cite{Luck,Dumont}. There, depending
on the actual substitution rule (and not on the substitution 
matrix alone), one can have the situation that $h(N)$ is bounded or
that $h(N)$ diverges logarithmically with $N$ (or, in other words,
$h_{n}$ diverges linearly with $n$). If $|\lambda^{(-)}|>1$, $h(N)$
can diverge like a power law. Again, polynomials reducible over the
integers are to be treated carefully, in particular for generalizations
to the $n$-letter case.

As an illustrative example, consider the substitution rules that 
have the substitution matrix
\be
\RRR_{\varrho} \;\; = \;\; 
\left( \ba{ll}
 2 & 1 \\
 1 & 2 \ea \right)
\ee
with eigenvalues $\lambda^{(+)}=3$ and $\lambda^{(-)}=1$, and
$p_{a}=p_{b}=1/2$. To obtain a unique limit word, we want
$w_{a}$ to commence with $a$, which leaves us with six different
substitution rules:
\be
\ba{l@{\hsp{50}}l@{\hsp{50}}l}
{\varrho_{1}} : \ba{ccc} a & \rightarrow & aab \\
                         b & \rightarrow & abb  \ea &
{\varrho_{2}} : \ba{ccc} a & \rightarrow & aab \\
                         b & \rightarrow & bab  \ea &
{\varrho_{3}} : \ba{ccc} a & \rightarrow & aab \\
                         b & \rightarrow & bba  \ea \\
{\varrho_{4}} : \ba{ccc} a & \rightarrow & aba \\
                         b & \rightarrow & abb  \ea & 
{\varrho_{5}} : \ba{ccc} a & \rightarrow & aba \\
                         b & \rightarrow & bab  \ea &
{\varrho_{6}} : \ba{ccc} a & \rightarrow & aba \\
                         b & \rightarrow & bba  \ea 
\ea\ee
Of these, $\varrho_{5}$ is special in the sense that it
leads to the periodic sequence $ababababa\ldots$ which
means that in this case the fluctuations are certainly
bounded (since in any finite part the numbers of $a$'s
and $b$'s differ at most by one). As it turns out, this
is only true for this special sequence, in all the five 
other cases $h_{n}$ grows linearly with $n$. More precisely,
one observes 
\be 
2 h_{n} \;\; = \;\; 
\left\{ \ba{l@{\hsp{20}}l}
n+1 & \mbox{for $\varrho_{1}$, $\varrho_{2}$, and $\varrho_{3}$} \\
n & \mbox{for $\varrho_{4}$} \\
1 & \mbox{for $\varrho_{5}$} \\
\max (1,n\! -\! 1) & \mbox{for $\varrho_{6}$} \ea \right.
\ee
for $n\geq 0$.

In Fig.~1, we show the different behaviour of the fluctuations
$g(N)$ for four typical substitution rules, namely
the Thue-Morse sequence (a),
the Silver Mean sequence (b) (sometimes also called ``Octonacci'' sequence),
the Period-Doubling sequence (c), and 
the Binary non-Pisot sequence (d). 
The substitution rules which define these sequences 
together with their statistical properties are 
summarized in Table~1. Fig.~2 shows the 
quantities $|g_{n}|$ (\ref{gN}) and $h_{n}$ (\ref{hN})
for the four different sequences. The marginal case of the
Period-Doubling sequence ($\lambda^{(-)}=-1$) clearly shows
the linear divergence of $h_{n}$ in $n$ whereas $|g_{n}|$
is constant.

%
%
%
%

\begin{center}
\vspace{5mm}
{\footnotesize
Table 1:\hsp{5} 
Four typical substitution rules and the statistical properties of
the corresponding sequences \\[3mm]
\begin{tabular}{|@{\hsp{8}}c@{\hsp{8}}|@{\hsp{8}}c@{\hsp{8}}
|@{\hsp{8}}c@{\hsp{8}}|@{\hsp{8}}c@{\hsp{8}}|@{\hsp{8}}c@{\hsp{8}}|}
\hline
sequence \rrr{6}{18} & substitution rule &
eigenvalues $\lambda^{(\pm)}$ & values of $p_{a}$ and $p_{b}$ &
$p_{a}/p_{b}$ \\
\hline
Thue-Morse \rrr{12}{30} & \subst{tm}{ab}{ba} & 
$2$, $\, 0$ & $\frac{1}{2}$, $\frac{1}{2}$ & $1$ \\
Silver Mean \rrr{12}{30} & \subst{sm}{aab}{a} & 
$1\pm\sqrt{2}$ & $1-\frac{\sqrt{2}}{2}$, $\frac{\sqrt{2}}{2}$ &
$\sqrt{2}-1$ \\
Period-Doubling \rrr{12}{30} & \subst{pd}{ab}{aa} & 
$2$, $-1$ & $\frac{2}{3}$, $\frac{1}{3}$ & $2$ \\
Binary non-Pisot \rrr{12}{30} & \subst{bnp}{ab}{aaa} & 
$\frac{1\pm\sqrt{13}}{2}$ & $\frac{5-\sqrt{13}}{2}$, 
$\frac{-3+\sqrt{13}}{2}$ & $\frac{1+\sqrt{13}}{2}$ \\
\hline
\end{tabular} }
\vspace{5mm}
\end{center}

%
%
%
%

\section{Critical Point and Fermion Velocity}
\setcounter{equation}{0}

We now come back to the Hamiltonian (\ref{mix}). We relate
the coupling constants $\varepsilon_{j}$ to a substitution
rule $\varrho$ (\ref{subs}) in the following way:
\be
\varepsilon_{j} \;\; = \;\; 
\left\{ \ba{l@{\hspace*{5mm}}l} 
\varepsilon_{a} & \mbox{if $j$-th letter in $w$ is $a$} \\
\varepsilon_{b} & \mbox{if $j$-th letter in $w$ is $b$.} \ea \;\right.
\ee
The condition $\mu=0$ (\ref{mu}) for criticality
now reads
\be 
p_{a} \log(\varepsilon_{a}) + p_{b} \log(\varepsilon_{b})
\;\; = \;\; 0
\ee
which has the following one-parameter solution \cite{Benza,RoBa}
\be
\varepsilon_{a} \;\; =\;\; r^{-p_{b}}\; , \hspace*{5mm}
\varepsilon_{b} \;\; =\;\; r^{p_{a}}\;  ,
\label{critcoup}
\ee
where $r>0$ is any positive real number.

As is well known (compare \cite{Cardy} and references therein), 
the finite-size scaling limit of the periodic
Ising quantum chain (which corresponds to $r\! =\! 1$) is described
by the $c\! =\! 1/2$ conformal field theory of a free Majorana fermion.
To obtain a conformally invariant scaling limit one has to have
a linear dispersion relation at criticality, which means that
the limit in Eq.~(\ref{v}) must exist. This in turn is only 
to be expected if the fluctuations $g(N)$ remain bounded, i.e.,
$h_{n}<S(\varrho)$ for all $n$, since at criticality one obtains from  
Eqs.~(\ref{gN}) and (\ref{critcoup}) 
\be
\prod_{j=1}^{N} \varepsilon_{j} \;\; = \;\; r^{-g(N)}
\ee
and all products of this type enter in Eq.~(\ref{v}).
On the other hand, if the limit in Eq.~(\ref{v}) does exist, 
the scaling limit will be the same as for the periodic 
quantum Ising chain.

Although we are not going to use it in this note,
we should add that there is a very elegant
and  efficient way to investigate this kind of
systems by means of the corresponding trace map
(compare \cite{tracemap,RoBa} and references therein). 
In this context, the critical point is obtained from a
unique one-parameter family of bounded orbits in the accessible
phase space region of the trace map \cite{RoBa}.

%
%
%
%

\section{Fermion and Conformal Spectrum}
\setcounter{equation}{0}

We now consider the spectrum of the Hamiltonian
(\ref{mix}) at criticality. It is described in terms of $N$
fermion frequencies $\Lambda_{k}\geq 0$. In Fig.~3, we present
the integrated density of the $\Lambda_{k}$ (which we
devided by the their largest value for convenience) for
coupling constants defined by $r=2$ (\ref{critcoup})
and several sizes of the chain.
In general there are no exact degeneracies in the spectrum
in contrast to the periodic case. However, as the size of the system
increases, the frequencies tend to accumulate which creates
the nearly vertical steps in Fig.~3, especially close
to the maximal frequency. 
The plots show characteristic gaps in the fermion
spectrum the locations of which (on vertical axis) are in accordance with 
the general gap labeling theorem \cite{Belli1a,Belli1b,tracemap}.
We should mention though that the Thue-Morse chain does not
show closed gaps here, in contrast to the situation
with electronic spectra of Schr\"odinger operators \cite{Belli2},
a phenomenon that deserves further exploration.  

For models whose continuum limit is described by a
conformal field theory, conformal invariance specifies
the behaviour of the low-energy excitations in the infinite
size limit \mbox{$N\rightarrow\infty$}. Essentially, they have to show
a leading $1/N$ behaviour and the 
level and degeneracy structure of the spectrum 
(after appropriate overall scaling of the gaps) 
is described by representations of the Virasoro algebra
with central extension $c$ which is the central charge of the
conformal field theory (see e.g.\ \cite{Cardy,Cardy2,Ginsparg}). 
This of course means that it is the {\em lower part}
of the integrated fermion density shown in Fig.~3 which is 
important for the conformal spectra.

To have a closer look at this part, we consider the
scaled energy gaps 
\be
\ol{E_{j}} \;\; = \;\; \frac{N}{2\pi v} \:\left(E_{j} - E_{0} \right)
\label{scaledgaps}
\ee
of the Hamiltonian $\tilde{H}^{(+)}$ (\ref{mix}), where $E_{0}$ denotes
the ground-state energy and $v$ is the velocity of elementary
excitations. 
For the Thue-Morse and the Silver Mean sequences, $v$ is
finite (in the limit $N\rightarrow\infty$ and for finite $r>0$) 
and given by \cite{Luck}
\be
v \;\; = \;\ \left\{ 
\ba{l@{\hsp{20}}l} 
{\left( \frac{r^{1/2} + r^{-1/2}}{2}\right)}^{-2} &
\mbox{for the Thue-Morse chain} \\
{\left(\frac{r-r^{-1}}{2}\right)}^{-1}\:\log (r) &
\mbox{for the Silver Mean chain.} \ea
\right.
\label{vv}
\ee

In fact, the second value generally applies for
all quasiperiodic sequences \cite{Luck} which can be obtained
from a certain section through a higher-dimensional periodic
structure via the dualization method \cite{Schlotty}.
On the other hand, the symmetric random dimer chain 
yields the same result as the Thue-Morse chain \cite{Luck}. 
It thus represents a disordered model (with bounded
fluctuations) which nevertheless
shows the same critical behaviour as the periodic
Ising chain and, in particular, leads to a conformally invariant
continuum limit.

{}From our discussion of the fluctuations and from Eq.~(\ref{v})
above it is clear that for the Period-Doubling and Binary non-Pisot
sequences the fermion velocity $v$ should vanish.
In Fig.~4, the scaled spectrum of low-energy 
excitations is shown for our four exemplary sequences.
For the Thue-Morse and Silver Mean chains, the
fermion velocity (\ref{vv}) is used whereas for the
other two sequences we simply normalize the first gap to $1/2$.
(We did not systematically investigate the scaling laws here,
although this might give independent access to the critical exponents
calculated in \cite{Luck}).
The formation of the so-called conformal towers can
be seen clearly for the spectra of the Thue-Morse and
Silver Mean chain (a look at the actual data confirms
that the degeneracies are those predicted
from conformal invariance),
whereas the spectra of the other two 
systems do not show any apparent regularities. As expected,
the normalization factors grow with the size of the system
in these cases. As a consequence, the nature of the critical
point is quite different, compare \cite{Luck}, and conformal
invariance is lost.

%
%
%
%

\section{Concluding Remarks}
\setcounter{equation}{0}

The interest in conformally invariant phase transitions 
and the almost immediate study of quasiperiodic systems 
of the Fibonacci type 
(see e.g.\ \cite{Benza,Ceccatto,YouZengXieYan,HenkelPatkos})
originally has led to the somewhat misleading conclusion 
that conformal invariance is robust with respect to many 
sorts of order or even disorder.
Although this is true for quasiperiodic Ising quantum chains 
(because that implies the PV-property or the bounded fluctuation 
property \cite{Bombi}), many other ordered structures can be 
defined which destroy conformal invariance at the critical point. 
Even more, one can consider the bounded fluctuation situation 
to be somewhat exceptional (e.g., PV-numbers are nowhere dense in 
$[1,\infty)$, see \cite{Boyd}) and thus conclude that the general 
case of a non-periodic Ising quantum chain 
does not lead to a conformally invariant scaling limit.

%
%
%
%

\section*{Acknowledgements}

It is a pleasure to thank Dieter Joseph for critically reading the manuscript.
M.\ B.\ would like to thank Paul A.\ Pearce and the Department of Mathematics,
University of Melbourne, for hospitality, where part of this work was done.
This work was supported by Deutsche Forschungsgemeinschaft.



%
%
%
%

\renewcommand{\baselinestretch}{1.1}

\clearpage
\begin{center}
{\footnotesize Figure 1:\hsp{5}
\parbox[t]{140mm}{Fluctuations $g(N)$, compare (\ref{gN}), for the following 
 sequences:  (a)~Thue-Morse, (b)~Silver Mean, \hsp{1}
 (c)~Period-Doubling, and \hsp{1} (d)~Binary non-Pisot. 
 The lengths of sequences which correspond to complete
 iteration steps are indicated.}\\[3mm] 
\mbox{\epsfbox{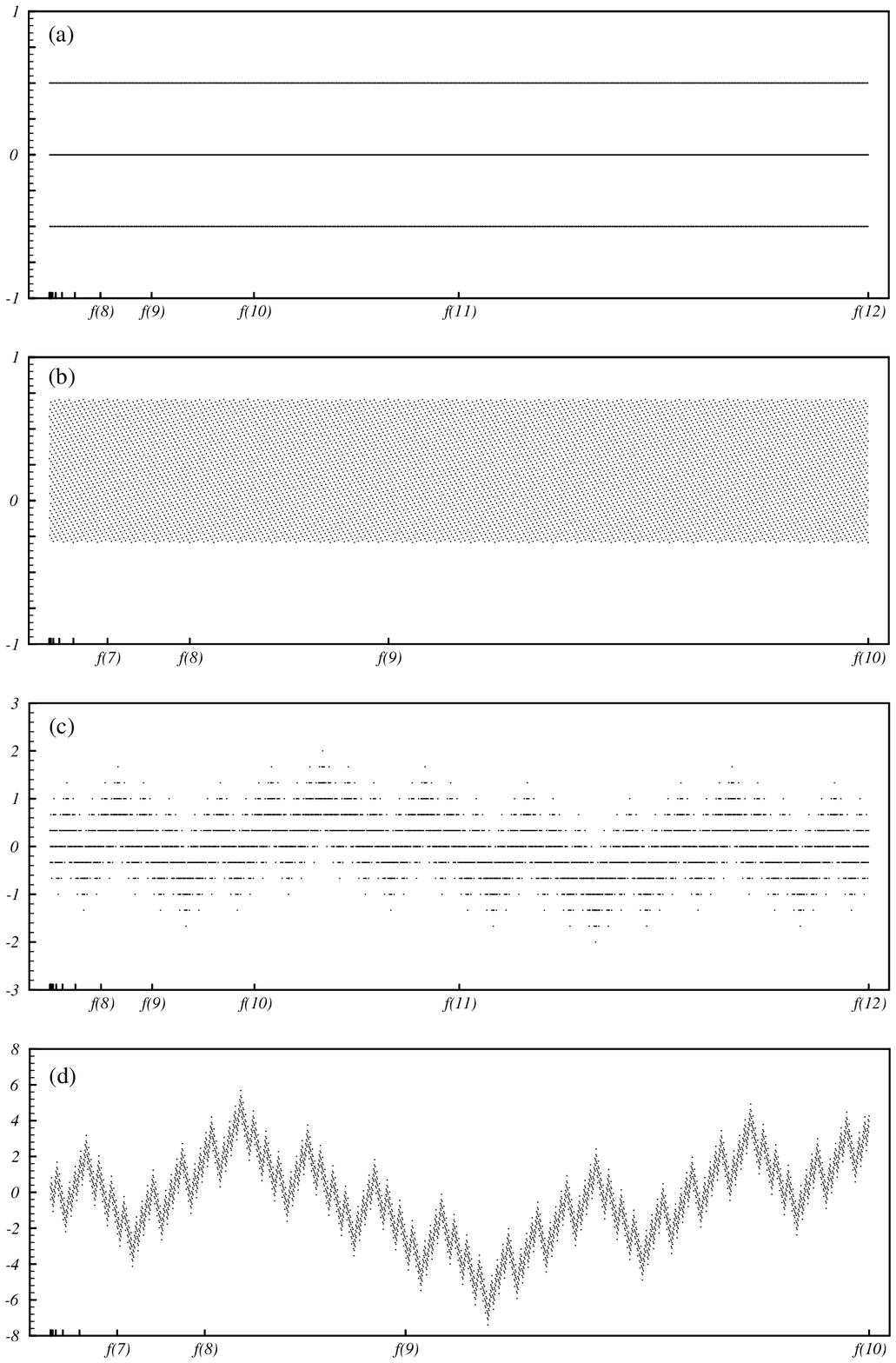}}
}
\end{center}

%
%
%
%
\clearpage
\begin{center}
{\footnotesize Figure 2:\hsp{5}
\parbox[t]{140mm}{Fluctuations $|g_{n}|$ (\ref{gN}) (Figure 2A) and 
$h_{n}$ (\ref{hN}) (Figure 2B) for the following 
 sequences: \hsp{1} (a)~Thue-Morse, \hsp{1} (b)~Silver Mean, \hsp{1}
 (c)~Period-Doubling, and \hsp{1} (d)~Binary non-Pisot.}\\[5mm]
\mbox{\epsfbox{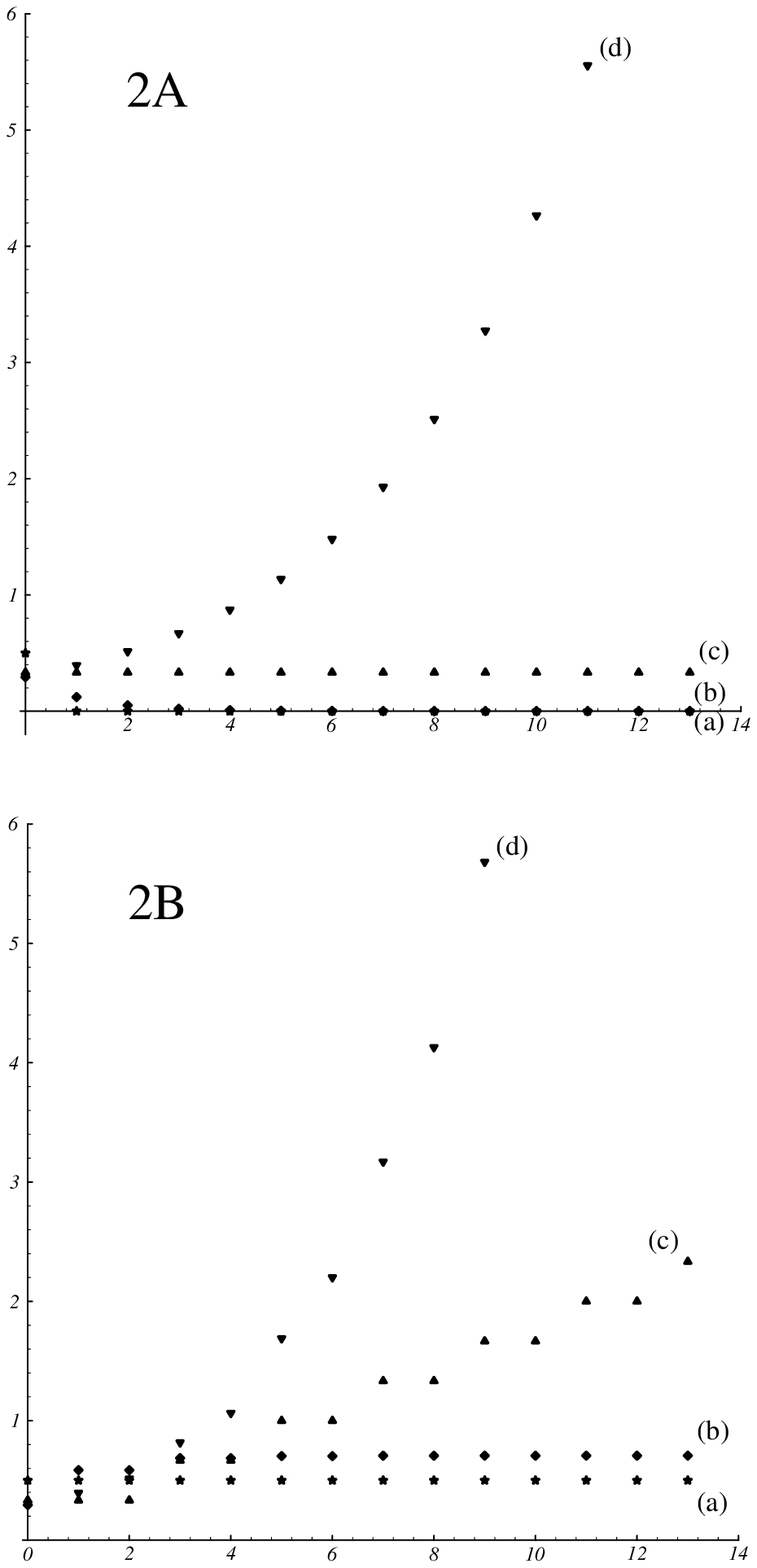}}
}
\end{center}

%
%
%
%
\clearpage
\begin{center}
{\footnotesize Figure 3:\hsp{5}
\parbox[t]{140mm}{Integrated density of 
 normalized fermion frequencies
 for the Hamiltonian  $\tilde{H}^{(+)}$ (\ref{mix}). The
 coupling constants are given by Eq.~(\ref{critcoup}) with $r=2$ and
 \hsp{1} (a1)--(a3) the Thue-Morse sequence with $n=5,6,7$,
 \hsp{1} (b1)--(b3) the Silver Mean sequence with $n=4,5,6$,
 \hsp{1} (c1)--(c3) the Period-Doubling sequence with $n=5,6,7$, and
 \hsp{1} (d1)--(d3) the Binary Non-Pisot sequence with $n=4,5,6$,
where $n$ denotes the number of iterations. The spectra do not
have true degeneracies, and the integrated density 
reaches 1 in all cases shown.}\\[5mm]
\mbox{\epsfbox{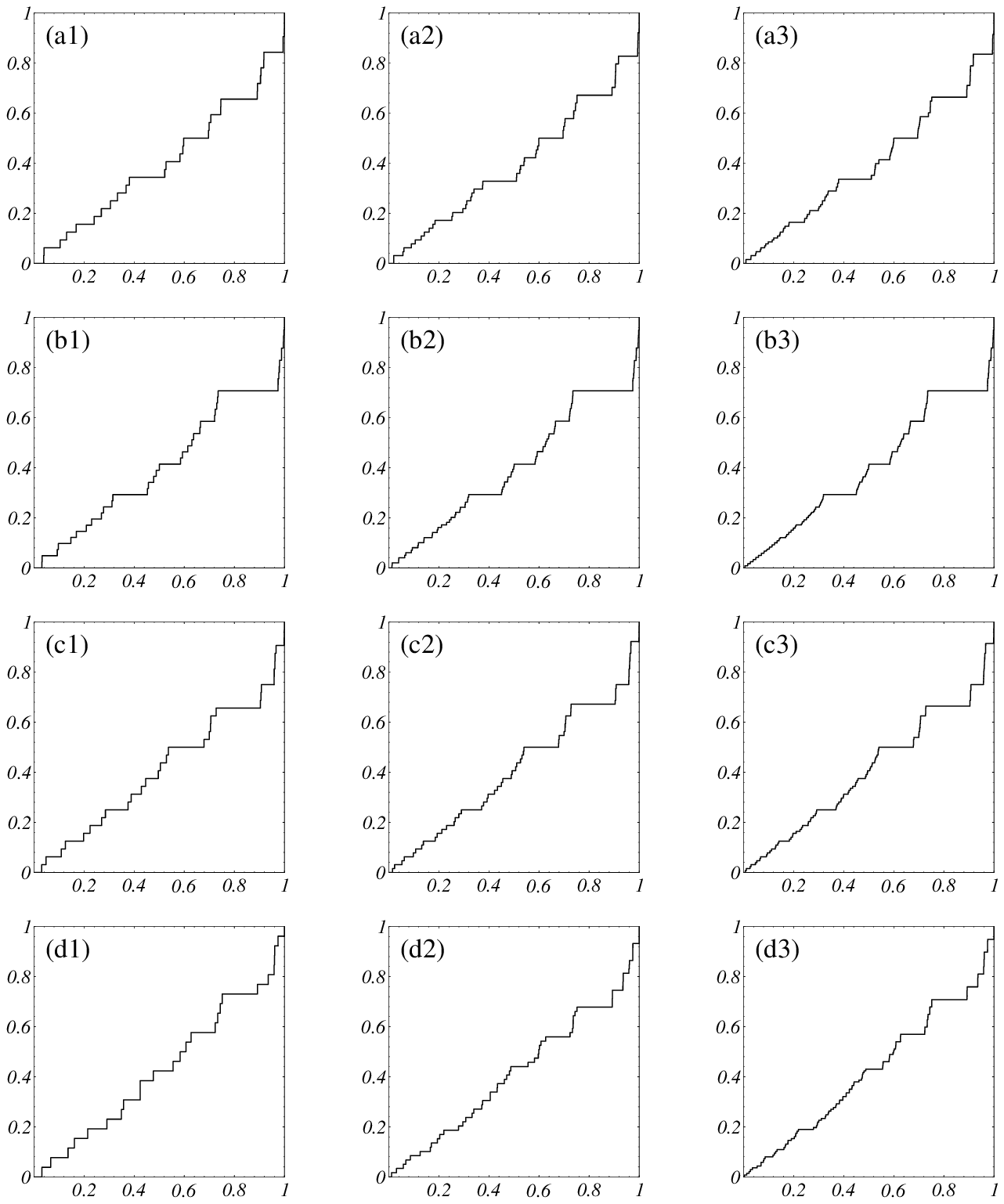}}
}
\end{center}

%
%
%
%
\clearpage
\begin{center}
{\footnotesize Figure 4:\hsp{5}
\parbox[t]{140mm}{Scaled low-energy spectra of $\tilde{H}^{(+)}$
 (\ref{mix}) for couplings obtained from Eq.~(\ref{critcoup}) with $r=2$ and
  (a1)--(a3) the Thue-Morse sequence with $n=6,7,8$,
 \hsp{1} (b1)--(b3) the Silver Mean sequence with $n=5,6,7$,
 \hsp{1} (c1)--(c3) the Period-Doubling sequence with $n=6,7,8$, and
 \hsp{1} (d1)--(d3) the Binary Non-Pisot sequence with $n=5,6,7$,
where $n$ denotes the number of iterations. Normalization of (a) and
(b) is taken from (\ref{vv}), while for (c) and (d) the first gap
is normalized to $1/2$.}\\[5mm]
\mbox{\epsfbox{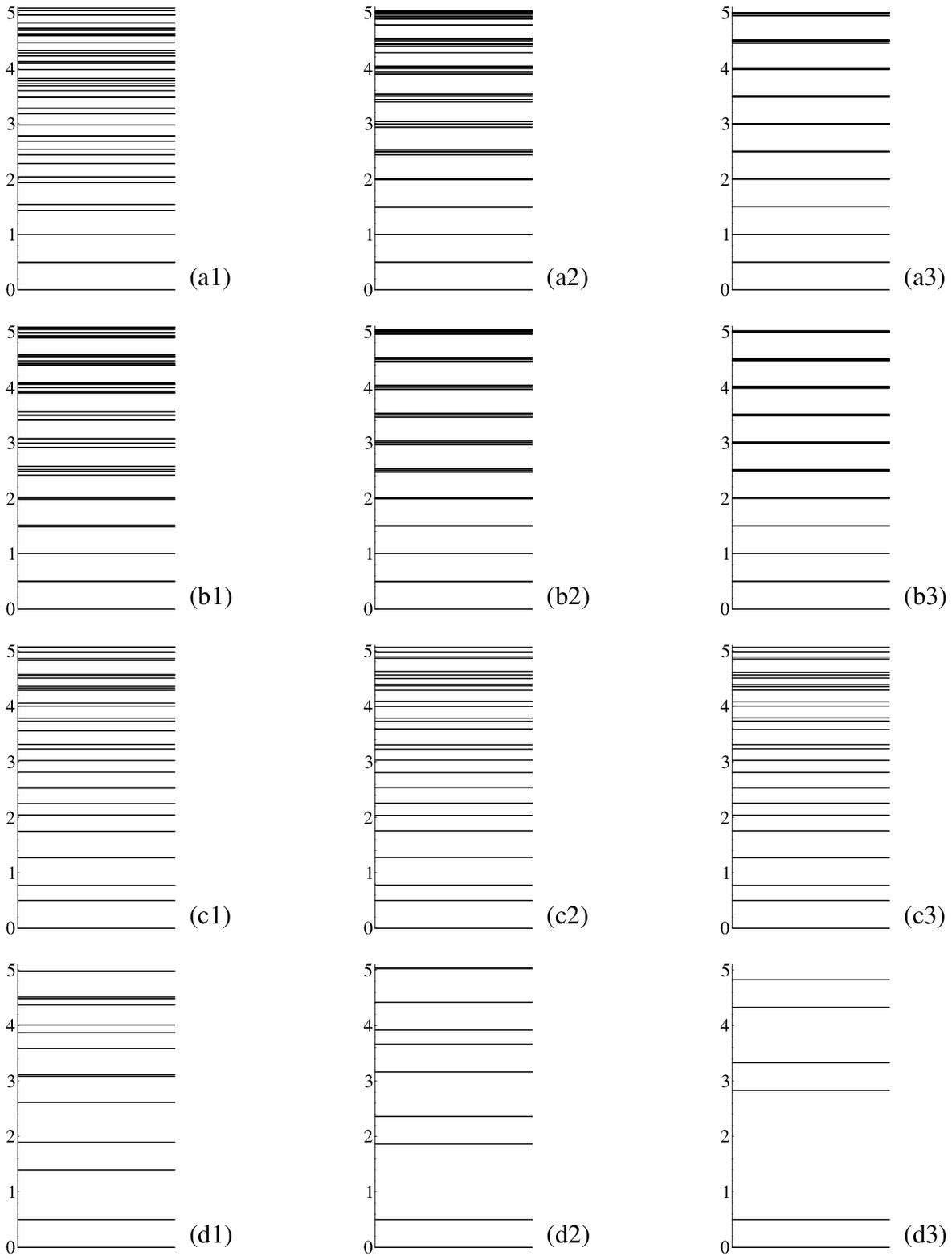}}
}
\end{center}

\end{document}